\documentclass[prl,aps,twocolumn,preprintnumbers,showpacs,superscriptaddress]{revtex4}

\usepackage{graphics,graphicx,epsfig,bm,amsthm,amsmath}

\begin{document}

\title{A Quantum-Quantum Metropolis Algorithm}

\date{\today}

\author{Man-Hong Yung}
\email{mhyung@chemistry.harvard.edu}
\affiliation{Department of Chemistry and Chemical Biology, Harvard University, Cambridge MA, USA}

\author{Al\'{a}n Aspuru-Guzik}
\email{aspuru@chemistry.harvard.edu}
\affiliation{Department of Chemistry and Chemical Biology, Harvard University, Cambridge MA, USA}


\begin{abstract}
Recently, the idea of classical Metropolis sampling through Markov chains has been generalized for quantum Hamiltonians. However, the underlying Markov chain of this algorithm is still classical in nature. Due to Szegedy's method, the Markov chains of classical Hamiltonians can achieve a quadratic quantum speedup in the eigenvalue gap of the corresponding transition matrix. A natural question to ask is whether Szegedy's quantum speedup is merely a consequence of employing classical Hamiltonians, where the eigenstates simply coincide with the computational basis, making cloning of the classical information possible. We solve this problem by introducing a quantum version of the method of Markov-chain quantization combined with the quantum simulated annealing (QSA) procedure, and describe explicitly a novel quantum Metropolis algorithm, which exhibits a quadratic quantum speedup in the eigenvalue gap of the corresponding Metropolis Markov chain for any quantum Hamiltonian. This result provides a complete generalization of the classical Metropolis method to the quantum domain. 
\end{abstract}

\maketitle
Interacting many-body problems, classical or quantum mechanical, generally require an exponentially large amount of computing resources to find the (exact) solutions, as the system size increases. Nonetheless, ingenious classical methods such as Markov-chain Monte Carlo (MCMC) or quantum Monte Carlo (QMC) have been invented and proven to be highly successful in many applications. These methods, however, have certain limitations. For example, the running time of MCMC scales as $O(1/\delta)$ \cite{Aldous1982}, where $\delta$ is the gap of the transition matrix. For problems such as spin glasses where $\delta$ is small, MCMC becomes computationally inefficient. QMC methods, on the other hand, suffer from the negative sign problem \cite{Troyer2005}. Despite many efforts have been made for improvement \cite{Kalos2000}, this limitation is still one of the biggest challenges in QMC \cite{Foulkes2001}.   

On the other hand, one of the most important goals in the field of quantum computation, as proposed by Feynman, is to look for new methods or algorithms that can solve these many-body problems more efficiently. This is referred to as {\it quantum simulation}, which can be implemented either by dedicated quantum simulators \cite{Buluta2009}, or with universal quantum computers \cite{Kassal2010}. For the former case, high precision experimental techniques are required for faithful simulation, which are therefore closer to engineering problems. For the latter case, sophisticated quantum algorithms are needed. To this end, there are two main approaches, (a) bottom-up approaches \cite{Zalka1998, Abrams1999, Lidar1999, Ortiz2001, Aspuru2005, Poulin2009}: to look for entirely new algorithms based on the special properties of quantum computers, and (b) top-down approaches \cite{Terhal2000, Szegedy04, Richter2007, Somma2008, Wocjan2008, Temme09, Yung2010}: to improve the existing classical algorithms by combining with elementary quantum algorithms. This work belongs to the latter class. 

Some years ago, Szegedy \cite{Szegedy04} described a method to quantize classical Markov chains. The key result is that a quadratic speedup $O(1/\sqrt{\delta})$ in the gap $\delta$ of the transition matrix is possible. This was later adapted to some new algorithms \cite{Somma2008, Wocjan2008}  that can prepare thermal states of classical Hamiltonians, based on the idea of quantum simulated annealing (QSA). 

In fact, Markov chains that correspond to classical Hamiltonians are relatively easy to construct, as all the eigenstates are in the computational basis $\left| x \right\rangle$. Preparation of the thermal states of quantum Hamiltonians \cite{Poulin2009,Terhal2000,Temme09}, however, is a much more challenging problem. It is because, classically, one needs to solve for the full eigenvalue problem (i.e., look for all eigenvalues and eigenvectors) for the quantum Hamiltonian first, which often requires more computational resources. 

The key to overcome this difficulty with a quantum computer is the use of quantum phase estimation algorithm \cite{Abrams1999}; the eigenvalues of a quantum Hamiltonian can be recorded without explicitly knowing the detailed structure of the eigenvectors. Based on this idea, Terhal and DiVincenzo \cite{Terhal2000} were able to extend the Metropolis algorithm to the quantum domain, avoiding the negative sign problem in QMC. However, one limitation for their results is that the Metropolis step involve too many energy non-local transitions, making it conceivably inefficient. Recently, Temme {\it et al}., \cite{Temme09} addressed this problem by introducing random local unitary operations in the update rule. In both cases, however, the underlying Markov chain is still classical in nature, which means that the scaling of the running time is still $O(1/\delta)$. 

\begingroup
\squeezetable
\begin{table}[t]
\caption{Comparison of various Markov-chain based algorithms for thermal state preparation}
\begin{center}
\begin{ruledtabular}
\begin{tabular}{l c l c c}
{\bf Methods} & {\bf Hamiltonian}\footnote{Quantum Hamiltonians include classical Hamiltonians.} & {\bf Input\footnote{Here, $\rho$ is a density matrix, $\left|  +  \right\rangle  \equiv \left( {\left| 0 \right\rangle  + \left| 1 \right\rangle } \right)/\sqrt 2$, and $\left| {\alpha ^0 } \right\rangle$ is defined in Eq. (\ref{int_state}).}} & {\bf Output\footnote{Here $\rho _{th}$ is the thermal density matrix (cf. Eq. (\ref{rho_th})), CETS II is defined in Eq. (\ref{coh_therm}), and CETS I (see also Ref. \cite{Yung2010}) is similar to CETS II, but the $\left\{ {\left| i \right\rangle } \right\}$ is replaced by the computational basis. Both of them are equivalent to $\rho_{th}$.}} & {\bf Q. Speedup\footnote{We consider only the quantum speedup with respect to the gap $\delta$ of the transition matrix of the Markov chain.}}\\
Q. Metropolis I\footnote{In {Ref.~\cite{Terhal2000}}, the Metropolis rule is implemented by a controlled-swap with a ``duplicated environment".} & Quantum & Any $\rho$ & $\rho_{th}$ & No\\
Q. Metropolis II\footnote{In Ref. \cite{Temme09}, improvement of Ref. \cite{Terhal2000}, a rejection rule for quantum states is imposed to implement the Metropolis rule.} & Quantum & Any $\rho$ & $\rho_{th}$ & No\\
Q. Markov chain I\footnote{In Ref. \cite{Somma2008}, combining Szegedy's Markov-chain quantization with quantum simulated annealing (QSA).} & Classical & $\left|  +  \right\rangle ^{ \otimes n}$ & CETS I & Quadratic\\
Q. Markov chain II\footnote{In Ref. \cite{Wocjan2008}, improvement of Ref. \cite{Somma2008} using Grover's fixed-point search.} & Classical & $\left|  +  \right\rangle ^{ \otimes n}$ & CETS I & Quadratic\\
$\rm Q^2MA$ (This work) & Quantum & $\left| {\alpha ^0 } \right\rangle$ & CETS II & Quadratic
\end{tabular}
\end{ruledtabular}
\end{center}
\label{table}
\end{table}%
\endgroup

In this letter, we propose a new quantum Metropolis algorithm for an arbitrary quantum Hamiltonian $H$ at a given temperature $T$ ($\beta \equiv 1 / k_B T$), with a quadratic speedup $O(1/\sqrt{\delta})$. We call this algorithm the quantum-quantum Metropolis algorithm ($\rm Q^2 MA$), as it shows a quantum speedup for the Markov chain of a quantum system being simulated. The previous version of the quantum metropolis algorithm \cite{Temme09} is restricted by the no-cloning theorem insofar that the information of a eigenstate cannot be retrieved after the proposed move in the Metropolis step. We relax this restriction by adopting a dual representation where the basis states consists of pairs of eigenstates related by time-reversal operation. For the cases where the physical system being simulated is time-reversal invariant, the pair of the  eigenstates forming the basis vector become identical. A comparison of various Markov-chain based methods is summarized in Table \ref{table}. 

The goal of the $\rm Q^2MA$ is to prepare the coherent encoding of the thermal state (CETS) $\left| {\alpha _0 } \right\rangle$ (cf. Eq. (\ref{coh_therm})), which, after tracing out the ancilla qubits, is equivalent to
\begin{equation}\label{rho_th}
\rho _{th}  = \frac{1}{Z}\sum\limits_i {e^{ - \beta E_i } \left| {\varphi _i } \right\rangle \left\langle {\varphi _i } \right|} \quad,
\end{equation}
where $Z = Tr\left[ {e^{ - \beta H} } \right]$ is the partition function, and ${\left| {\varphi _i } \right\rangle }$ is the eigenstate of a quantum Hamiltonian $H$, associated with the eigenvalue $E_i$. Before going into the technical details of this work, we briefly summarize the important features of Markov chains and Metropolis sampling, in a way applicable to both classical and quantum systems.
 
\emph{Brief review of the Metropolis method ---}  In the standard Metropolis scheme, the Gibbs distribution $e^{-\beta E_i}/Z$ of certain eigenstates is generated through a Markov chain $M$, in which the matrix element $m_{ij}$ refers to the transition probability from the eigenstate ${\left| {\varphi _i } \right\rangle }$ to ${\left| {\varphi _j } \right\rangle }$. The equilibrium (stationary) distribution 
 \begin{equation}
 \pi_i \equiv e^{-\beta E_i}/Z 
\end{equation} 
 satisfies the detailed balance condition,
 \begin{equation}\label{de_bal_con}
\pi _i m_{ij}  = \pi _j m_{ji} \quad.
\end{equation}
A solution which can satisfy the detailed balance condition is $m_{ji} = s_{ij} z_{ji}$, where $s_{ij}=s_{ji}$ is any symmetrical transition probability, 
\begin{equation}\label{z_ij}
z_{ij}  = \min \{ {1,e^{ - \beta \left( {E_j  - E_i } \right)} } \}
\end{equation}
is sometimes called the Metropolis filter. In the practical implementation of the Metropolis method, one starts with some initial configuration, and then apply a random transition (e.g. single spin-flip) and compare the energy between the new eigenstate and the old one. If the new eigenstate has a lower energy, accept the move. Otherwise, accept the move only with a probability distribution given by the ratio of the corresponding Boltzmann factors $e^{ - \beta \left( {E_j  - E_i } \right)}$. This is called the Metropolis method.

The performance of the Metropolis method depends on the properties of the underlying transition matrix $M$ of the Markov chain, especially, the spectrum of the eigenvalues $\lambda_k$, which are all positive and bounded by $1$, and the largest eigenvalue is always $1$. For convenience, we order them as  
\begin{equation}\label{lambda_sequence}
\lambda_0  = 1 > \lambda_1  \ge ... \ge \lambda_{N - 1}  > 0 \quad.
\end{equation}
The convergence time of a Markov chain is limited by the eigenvalue gap $\delta \equiv 1 - \lambda_1$ of the transition matrix $M$ by $O(1/\delta)$ \cite{Aldous1982}.  The purposes of this work are: (1) to show that the running time can be improved to $O(1/\sqrt{\delta})$, and (2) to fully extend the Metropolis sampling algorithm into the quantum domain.

\emph{Generalization of the Markov-chain quantization ---} The original Markov-chain quantization method by Szegedy \cite{Szegedy04} is applicable to classical Hamiltonians only. To extend it to quantum Hamiltonians, we start with $n$ qubits prepared in the state $\left( {1/\sqrt 2 } \right)^n \left( {\left| 0 \right\rangle  + \left| 1 \right\rangle } \right)^{ \otimes n}$, or equivalently $N^{ - 1/2} \sum\nolimits_{x = 0}^{N - 1} {\left| x \right\rangle }$, where $N=2^n$. By performing a bit-by-bit CNOT gate on a set of $n$ ancilla qubits initialized in the state $\left| {000...0} \right\rangle$, one can {\it formally} express the entire quantum state as 
\begin{equation}\label{int_state}
\left| {\alpha^0 } \right\rangle  \equiv \frac{1}{{\sqrt N }}\sum\limits_{i = 0}^{N - 1} {\left| {\varphi _i } \right\rangle \left| {\tilde \varphi _i } \right\rangle }   \quad,
\end{equation}
where $\left| {\varphi _i } \right\rangle  = \sum\nolimits_{x = 0}^{N - 1} {\langle x \left| {\varphi _i } \right\rangle \left| x \right\rangle }$ is an energy eigenstate of $H$ (i.e., ${H\left| {\varphi _i } \right\rangle  = E_i \left| {\varphi _i } \right\rangle}$), and 
\begin{equation}
\left| {\tilde \varphi _i } \right\rangle  \equiv \sum_{x = 0}^{N - 1} {\langle {\varphi _i } | x \rangle \left| x \right\rangle }
\end{equation}
is the {\it time-reversal} counterpart of $\left| {\varphi _i } \right\rangle$, which is the eigenstate of the corresponding time-reversal Hamiltonian $\tilde H \equiv H ^*$ with the same eigen-energy $E_i$ (i.e., $\tilde H\left| {\tilde \varphi _i } \right\rangle  = E_i \left| {\tilde \varphi _i } \right\rangle$). Through the phase estimation algorithm (PEA), the value of eigenvalue $E_i$ can be obtained either from $\left| {\varphi _i } \right\rangle$ or $\left| {\tilde \varphi _i } \right\rangle$. This is the key property introduced in this paper which relaxes the constraints of the previous quantum Metropolis algorithm \cite{Temme09}.  In the following, for the purpose of demonstration, we shall assume that the PEA can be applied perfectly, in the sense that each eigenstate can be uniquely identified by a unique eigenvalue. We leave our discussion on the effects of degeneracy on this algorithm in the Appendix. It turns out that the degeneracy of the eigen-energies of a quantum Hamiltonian alone is not necessarily a problem. 

Now, let us include into our Hilbert space an extra qubit initialized in $\left| 0 \right\rangle$ and define a more compact notation, 
\begin{equation}
\left| {i} \right\rangle  \equiv \left| {\varphi _i } \right\rangle \left| {\tilde \varphi _i } \right\rangle \left| 0 \right\rangle \quad.
\end{equation}
The information of a Markov chain can be encoded in a pair of unitary operators $U_X$ and $U_Y$ (see Appendix for their detailed construction),
\begin{eqnarray}
U_X \left| {i } \right\rangle  {=} \sum\limits_k {\left( {\sigma _{ik} \left| {\varphi _i } \right\rangle \left| {\varphi _k } \right\rangle \left| 0 \right\rangle  + \gamma _{ik} \left| {\varphi _i } \right\rangle \left| {\varphi _k } \right\rangle \left| 1 \right\rangle } \right)}   \, , \label{def_U_X} \\ 
 U_Y | {j} \rangle  {=} \sum\limits_m {( {\sigma _{jm} | {\varphi _m } \rangle | {\varphi _j } \rangle | 0 \rangle  + \gamma _{jm} | {\varphi _j } \rangle | {\varphi _m } \rangle | 1\rangle } ) \, , \label{def_U_Y}} 
\end{eqnarray}
where $\sigma _{ik}  \equiv \alpha _{k\tilde i} \sqrt {z_{ik} }$, $\gamma _{ik}  \equiv \alpha _{k\tilde i} \sqrt {1 - z_{ik} }$, and $\alpha _{k\tilde i}  \equiv \left\langle {\varphi _k } \right|K\left| {\tilde \varphi _i } \right\rangle$. Here $K$ is an unitary operator which plays the same role as the spin-flip in the classical Metropolis method, and $z_{ik}$ is the Metropolis filter defined in Eq. (\ref{z_ij}). Note that $U_Y$ is related to $U_X$ by a controlled-SWAP operation.

\emph{Satisfying the detailed balance conditions ---} 
The detailed balance condition (Eq. (\ref{de_bal_con})) can be recovered by the product of $U_X^\dagger$ and $U_Y$. From Eq. (\ref{def_U_X}) and (\ref{def_U_Y}), for $j \ne i$, we have
\begin{equation}\label{U_i_to_j}
\left\langle {j } \right|U_X^\dagger  U_Y \left| {i} \right\rangle  = | { {\alpha _{j\tilde i} }  } |^2 ({z_{ij} z_{ji} })^{1/2}  \quad,
\end{equation}
where we used $\langle {\varphi _i } | {\tilde \varphi _j } \rangle  = \langle {\varphi _j } | {\tilde \varphi _i } \rangle$. On the other hand, 
\begin{equation}\label{U_i_to_i}
\langle {i } |U_X^\dagger  U_Y | {i } \rangle  {=} | {\alpha _{i\tilde i} } |^2  {+} \sum\limits_k {| {\alpha _{k\tilde i} } |^2 } ( {1 - z_{ik} } ),
\end{equation}
which, as we shall see, can be interpreted as the probability of not undergoing a transition. 

\emph{Construction of the operator $W$ ---}  Now, using Eq. (\ref{de_bal_con}), (\ref{U_i_to_j}) and (\ref{U_i_to_i}), we obtain the following decomposition:
\begin{equation}\label{decomp}
\left\langle j \right|U_X^\dagger U_Y \left| i \right\rangle  = \left\langle j \right|D_\pi ^{1/2} \left| j \right\rangle \left\langle j \right|M\left| i \right\rangle \left\langle i \right|D_\pi ^{ - 1/2} \left| i \right\rangle \quad,
\end{equation}
where $D_\pi   \equiv \sum\nolimits_{j = 0}^{N - 1} {\pi _j \left| j \right\rangle \left\langle j \right|}$ is a diagonal matrix, and $
M \equiv \sum\nolimits_{i,j} {m_{ij} \left| j \right\rangle \left\langle i \right|}$, with $m_{ii}  \equiv \left\langle i \right|U_X^\dagger  U_Y \left| i \right\rangle$ and $m_{ij}  \equiv | {\alpha _{j\tilde i} } |^2 z_{ij}$ for $j \ne i$ is the transition matrix of the Markov chain. Within the subspace $\left\{ {\left| i \right\rangle } \right\}$, Eq. (\ref{decomp}) implies that $U_X^\dagger  U_Y$ and $M$ are similar matrices, which means that they have the same set of eigenvalues $\lambda_k$ (see Eq. (\ref{lambda_sequence})). Following Szegedy \cite{Szegedy04}, this property allows us to construct an operator
\begin{equation}\label{Szegedy_W}
W \equiv \left( {2\Lambda _2  - I} \right)\left( {2\Lambda _1  - I} \right) \quad,
\end{equation}
where 
\begin{equation}\label{lambda1_2}
\Lambda _1  \equiv \sum\limits_{i = 0}^{N - 1} {\left| i \right\rangle \left\langle i \right|} \quad {\rm and} \quad \Lambda _2  \equiv U_X^\dagger  U_Y \Lambda _1 U_Y^\dagger  U_X \, .
\end{equation}
The spectral properties of $W$ can be seen in the following way: define $\left| {\alpha _k } \right\rangle  \equiv \sum\nolimits_{i = 0}^{N - 1} {a_{ki} \left| i \right\rangle } $ to be the eigenvectors of $\Lambda_1 U_X^{\dagger} U_Y \Lambda_1$, the eigenvalue equation can be written as
\begin{equation}\label{Lambda_1}
\Lambda _1 U_X^\dagger  U_Y \left| {\alpha _k } \right\rangle  = \lambda_k \left| {\alpha _k } \right\rangle \quad.
\end{equation}
On the other hand, using the fact that $\Lambda _1 U_X^\dagger U_Y \Lambda _1  = \Lambda _1 U_Y^\dagger  U_X \Lambda _1$, 
we have,
\begin{equation}\label{Lambda_2}
\Lambda _2 \left| {\alpha _k } \right\rangle  = \lambda_k U_X^\dagger  U_Y \left| {\alpha _k } \right\rangle \quad.
\end{equation}
Eq. (\ref{Lambda_1}) and (\ref{Lambda_2}) suggest that, if we start with vectors within $\Lambda_1$, $W$ can be block-diagonalized into subspace of $2 \times 2$ matrices $w_k$ spanned by the basis $\{ {\left| {\alpha _k } \right\rangle ,U_X^\dagger  U_Y \left| {\alpha _k } \right\rangle }\}$. Explicitly, 
\begin{equation}
w_k = \left[ {\begin{array}{*{20}c}
   {\cos \left( {2\theta _k } \right)} & { - \sin \left( {2\theta _k } \right)}  \\
   {\sin \left( {2\theta _k } \right)} & {\cos \left( {2\theta _k } \right)}  \\
\end{array}} \right] \quad,
\end{equation}
where $\cos \theta _k  \equiv \lambda_k$. Note that the eigenvalues of $w_k$ is $e^{ \pm i\theta _k }$. In the case of $k=0$ where $\lambda_0  = 1$ (or $\theta _0  = 0$), $w_0=I$ is simply an identity. From Eq. (\ref{decomp}) and the properties of the Markov matrix (see Appendix), the $k=0$ state is the coherent encoding of the thermal state (CETS) of Eq. (\ref{rho_th}),
\begin{equation}\label{coh_therm}
\left| {\alpha _0 } \right\rangle  = \sum\limits_{i = 0}^{N - 1} {\sqrt {\pi _i } } \left| i \right\rangle \quad.
\end{equation}
Recall that $\left| i \right\rangle  \equiv \left| {\varphi _i } \right\rangle \left| {\tilde \varphi _i } \right\rangle \left| {0 } \right\rangle$, the state $\left| {\alpha _0 } \right\rangle$ becomes the Gibbs thermal state  $\rho_{th}  = e^{ - \beta H} /Tr [{e^{ - \beta H} }]$ when the other qubits are traced out. 

One of the most important features about $W$ is that the minimum eigenvalue gap $\Delta _{\min }  \equiv \left|  2 \theta_1 \right|$ of $W$ is less than two times the square root of the gap $\delta  \equiv 1 - \lambda_1$ of the transition matrix $M$ (using $2\theta  \ge \left| {1 - e^{2i\theta } } \right| = 2\sqrt {1 - \cos ^2 \theta }$):
\begin{equation}\label{Delta_min}
\Delta _{\min }  \ge 2\sqrt \delta  \quad,
\end{equation}
which is the origin of the quadratic speedup of Szegedy's algorithm. This completes our discussion on the necessary tools needed for the following discussion. 

\emph{Quantum simulated annealing (QSA) ---}  Given a quantum Hamiltonian $H$ and any finite temperature $T$, our goal is to obtain the corresponding coherent thermal state of the form in Eq. (\ref{coh_therm}), from the initial state defined in Eq. (\ref{int_state}), which can be readily prepared from the ``all-zero" state $\left| {000...0} \right\rangle$, and be considered as the infinite-temperature state. To achieve this goal, we can use the method of quantum simulated annealing (QSA) \cite{Somma2008}. For completeness, we outline the basic strategy and summarize the related results in the following paragraph.

The strategy of the QSA method is to prepare a sequence, $j = 0,1,2,..,d$, of $d+1$ of coherent thermal states,
\begin{equation}\label{CETS_j}
| {\alpha^j_0 } \rangle  = \sum_{i = 0}^{N - 1} {\left( {e^{ - \beta _j E_i } /Z_j} \right)^{1/2} } \left| i \right\rangle \quad,
\end{equation}
at a time; the temperature $\beta _j  \equiv \left( {j/d} \right)\beta$ of the coherent thermal state is lowered in each step. The basic idea is that, for sufficiently small $\Delta \beta \equiv \beta /d$,  one can show that $| {\alpha_0^j }\rangle$ has a good overlap with $| {\alpha_0^{j + 1} }\rangle$, i.e., 
\begin{equation}
| {\langle {\alpha _0^{j + 1} } | {\alpha _0^j } \rangle } |^2  \ge 1 - \epsilon _0 \quad,
\end{equation}
where $\epsilon _0$ is bounded by  $O( {\Delta \beta ^2 \left\| H \right\|^2 } )$, then a projective measurement 
\begin{equation}
\Pi _{k + 1}  = \sum_{k = 0}^{N - 1} {| {\alpha _k^{j + 1} } \rangle } \langle {\alpha _k^{j + 1} } |
\end{equation}
on the eigenbasis of $| {\alpha _k^{j+1} } \rangle$ to $| {\alpha _0^{j} } \rangle$ will result in the lower temperature state $| {\alpha _0^{j + 1} } \rangle$ with a high probability. For the whole process, 
\begin{equation}
\left| {\alpha _0^0 } \right\rangle \buildrel {\Pi _1 } \over
 \longrightarrow \left| {\alpha _0^1 } \right\rangle \buildrel {\Pi _2 } \over
 \longrightarrow ...\buildrel {\Pi _d } \over
 \longrightarrow \left| {\alpha _0^d } \right\rangle \quad ,
\end{equation}
the total error (see Appendix)
\begin{equation}
\epsilon  = d\epsilon _0 < O( {\beta ^2 \left\| H \right\|^2 /d}) 
\end{equation}
is suppressed by refining the step size, analogous to the quantum Zeno effect. 

Now, using the machinery we have developed, the operator $W_{j+1}$  (the $W$ operator defined in Eq. (\ref{Szegedy_W}) for $| {\alpha _k^{j + 1} } \rangle$) can be used to construct such a projective measurement, through the phase estimation algorithm \cite{Wocjan2008} (see also Appendix).  Alternatively, one may perform an artificial way of introducing decoherence \cite{Somma2008}, where $W_{j+1}$ is applied multiple times randomly. In any case, the number of controlled-$W_{j+1}$ is at most $O(1/\sqrt{\delta})$, which is a quadratic speedup relative to classical Markov chains. This completes our description on the $\rm Q^2 MA$.

\emph{Conclusion ---} To summarize, we have described a new quantum Metropolis algorithm which extends Szegedy's method of classical Markov-chain quantization to the quantum domain, and provides a quadratic quantum speedup $O(1/{\sqrt {\delta}})$ in the gap $\delta$ of the transition matrix $M$. The restriction encountered by the previous version of the quantum Metropolis algorithm is mostly due to the no-cloning theorem, where the required information, such as the associated eigenvalue, of an eigenstate cannot be retrieved after the proposed move in the Metropolis step. We relax this restriction by adopting a dual representation where the set of basis states consists of pairs of eigenstates related by the time-reversal operation. 

This result completes the generalization of the classical Metropolis method to the quantum domain. Morevoer, the advantages of this quantum algorithm over classical algorithms could be exponential, as there is no need to explicitly solve for the eigenvalues and eigenvectors in the classical ways for the quantum Hamiltonians being simulated. Finally, as the application of the Metropolis method to quantum Hamiltonians can be considered as a special case of quantum maps (operations), it may be possible that the results presented here could be generalized to allow quantum speedup for a much broader class of quantum maps.



\begin{acknowledgments}
We thank S. Boixo, R. D. Somma, K. Temme, J. D. Whitfield and P. Wocjan for insightful discussions, and are grateful to the following funding sources: Croucher Foundation for M.H.Y; DARPA under the Young Faculty Award N66001-09-1-2101-DOD35CAP, the Camille and Henry Dreyfus Foundation, and the Sloan Foundation;  Army Research Office under Contract No. W911NF-07-1-0304 for A.A.G.
\end{acknowledgments}

\bibliographystyle{apsrev}

\appendix

\section{Appendix: Construction of the unitary operators $U_X$ and $U_Y$}
Here we show how one may construct the unitary operator $U_X$ defined in Eq. (\ref{def_U_X}), namely
\begin{equation}
U_X \left| {i } \right\rangle  {=} \sum\limits_k {\left( {\sigma _{ik} \left| {\varphi _i } \right\rangle \left| {\varphi _k } \right\rangle \left| 0 \right\rangle  + \gamma _{ik} \left| {\varphi _i } \right\rangle \left| {\varphi _k } \right\rangle \left| 1 \right\rangle } \right)}   \quad,
\end{equation}
where 
\begin{equation}
\sigma _{ik}  \equiv \alpha _{k\tilde i} \sqrt {z_{ik} } \quad {,} \quad \gamma _{ik}  \equiv \alpha _{k\tilde i} \sqrt {1 - z_{ik} } \quad,
\end{equation}
and 
\begin{equation}
\alpha _{k\tilde i}  \equiv \left\langle {\varphi _k } \right|K\left| {\tilde \varphi _i } \right\rangle \quad .
\end{equation}
Here $K$ is an unitary operator which plays the same role as the spin-flip in the classical Metropolis method, and $z_{ik}$ is the Metropolis filter defined in Eq. (\ref{z_ij}). Note that $U_Y$ is related to $U_X$ by a controlled-SWAP. To ensure Eq. (\ref{U_i_to_j}) is positive, we assume that $K$ is symmetrical in the computational basis:
\begin{equation}
\left\langle {x'} \right|K\left| x \right\rangle  = \left\langle x \right|K\left| {x'} \right\rangle  \quad.
\end{equation}
For example, $K$ can be the SWAP operation.

We start with the following $n$-qubit state
\begin{equation}
\left( {\frac{1}{{\sqrt 2 }}} \right)^n \left( {\left| 0 \right\rangle  + \left| 1 \right\rangle } \right)^{ \otimes n} \quad,
\end{equation}
which is equivalent to the ``all-input" state
\begin{equation}
\frac{1}{\sqrt{N}}\sum\limits_{x = 0}^{N - 1} {\left| x \right\rangle } \quad,
\end{equation}
where $N \equiv 2^n$. Suppose now we include a set of $n$ ancilla qubits initialized in the ``all-down" state 
\begin{equation}
\left| {000...0} \right\rangle \quad ,
\end{equation}
an apply a bit-by-bit CNOT operation, which is equivalent to a copy of the value of $x$ to the register qubits (which is not the same as quantum state cloning),
\begin{equation}
\left| x \right\rangle \left| {000...0} \right\rangle  \to \left| x \right\rangle \left| x \right\rangle \quad.
\end{equation}
The resulting state is 
\begin{equation}\label{app_resulting_state}
\frac{1}{{\sqrt N }}\sum\limits_{x = 0}^{N - 1} {\left| x \right\rangle } \left| x \right\rangle \quad.
\end{equation}

Given any Hamiltonian $H$, if we formally insert the completeness relation 
\begin{equation}
I = \sum_{i = 0}^{N - 1} {\left| {\varphi _i } \right\rangle \left\langle {\varphi _i } \right|} \quad,
\end{equation}
expanded in the eigenvector basis $\left\{ {\left| {\varphi _i } \right\rangle } \right\}$, to the state in Eq. (\ref{app_resulting_state}), then we get the state in Eq. (\ref{int_state}),
\begin{equation}
\left| {\alpha^0 } \right\rangle  \equiv \frac{1}{{\sqrt N }}\sum\limits_{i = 0}^{N - 1} {\left| {\varphi _i } \right\rangle \left| {\tilde \varphi _i } \right\rangle }   \quad,
\end{equation}
without solving the eigenvalue equation, where 
\begin{equation}
\left| {\tilde \varphi _i } \right\rangle  \equiv \sum_{x = 0}^{N - 1} {\langle {\varphi _i } | x \rangle \left| x \right\rangle }
\end{equation}
is the time-reversal counterpart of ${\left| {\varphi _i } \right\rangle }$. This state can be considered as the infinite-temperature state, and is also the starting point for the quantum simulated annealing (QSA). 

We are now ready to consider the explicit procedure for constructing $U_X$ defined Eq. (\ref{def_U_X}). Starting with the paired state 
\begin{equation}
\left| {\varphi _i } \right\rangle \left| {\tilde \varphi _i } \right\rangle \quad ,
\end{equation}
we apply the ``kick" operator $K$ to $\left| {\tilde \varphi _i } \right\rangle$, and write
\begin{equation}
K\left| {\tilde \varphi _i } \right\rangle  = \sum\limits_k {\alpha _{k\tilde i} \left| {\varphi _k } \right\rangle } \quad .
\end{equation}
Next, we implement the Metropolis filter by performing a controlled-rotation (based on the difference of the eigenvalues):
\begin{equation}
\left| {\varphi _i } \right\rangle \left| {\varphi _k } \right\rangle \left| 0 \right\rangle  \to \left| {\varphi _i } \right\rangle \left| {\varphi _k } \right\rangle \left( {\sqrt {z_{ik} } \left| 0 \right\rangle  + \sqrt {1 - z_{ik} } \left| 1 \right\rangle } \right) \, ,
\end{equation}
where 
\begin{equation}
z_{ij}  \equiv \left\{ {1,e^{ - \beta \left( {E_j  - E_i } \right)} } \right\} \quad .
\end{equation}
This creates a state as described by Eq. (\ref{def_U_X}). Note that only the information about the difference of the eigenvalues is needed. There is no need to determine each eigenvalue individually.

\section{Appendix: Generalization to include random ``kicks"}
In applying the classical Metropolis method, for example to Ising model, one usual apply random spin-flips to the spins. This feature can be incorporated in our $\rm Q^2MA$. On the other hand, implementation of these spin-flips is also necessary for systems with time-reversal symmetry where the eigenstates $\left| {\tilde \varphi _i } \right\rangle  = \left| {\varphi _i } \right\rangle$ contain only real coefficients, in the computational basis. 

To be specific, we consider a system of $n$ spin-1/2 particles, we include an extra ancilla qubits initialized as $\left| {000...0} \right\rangle$. Then, in the first step of $U_X$, we perform a transformation such that 
\begin{equation}
\left| {000...0} \right\rangle  \to {1 \over {\sqrt n }}\sum\limits_{\lambda =1}^{n} \left| \lambda  \right\rangle. 
\end{equation}
Then, conditioned on each value of $\lambda$, we apply a ``kick" operator $K_{\lambda}$, where $K_\lambda ^ \dagger   = K_\lambda$, e.g. spin-flip operator, to the spin-$\lambda$. The net effect is that Eq. (\ref{def_U_X}) becomes
\begin{equation}
\left\langle {j_0 } \right|U_X^ \dagger  U_Y \left| {i_0 } \right\rangle  = {1 \over n}\sum\limits_\lambda  {\left| {\left\langle {\varphi _j } \right|K_\lambda  \left| {\tilde \varphi _i } \right\rangle } \right|^2 \sqrt {z_{ij} z_{ji} } } \quad,
\end{equation}
and Eq. (\ref{def_U_Y}) changes in a similar way.

\section{Appendix: Effects of degeneracy and the limitations of the algorithm}
The $\rm Q^2MA$ presented in the main text does not necessarily break down when the Hamiltonian $H$ of the quantum system is highly degenerate in the space of the eigenstates. For the sake of the argument, consider the Ising model:
\begin{equation}
H_{\rm Ising} = J \sum\limits_{\left\langle {i,j} \right\rangle } {\sigma _i^z } \sigma _j^z  \quad.
\end{equation}
Although the eigenstates are highly degenerate, the $\rm Q^2MA$ does not require the knowledge of the individual eigenvalues, but instead, it needs the difference between two eigenstates before and after the ``kick" (spin-flip). In the case of the Ising model, the change in energy is $O(J)$, thus it is sufficient to ensure our resolution in resolving the energy change be smaller than $O(J)$.

To understand this point better, note that there are two places degeneracy would affect the agrument: (A) the implementation of the projector 
\begin{equation}
\Lambda _1  \equiv \sum\limits_i {\left| i \right\rangle \left\langle i \right|}
\end{equation}
in Eq. (\ref{lambda1_2}). (B) The ``leakage" to the degenerate subspace in Eq. (\ref{def_U_X}). These two points are related. The projector $\Lambda_1$ can be implemented via a filter method (e.g. see Ref. \cite{Wocjan2008}); this is essentially the same as applying the phase estimation algorithm (PEA) multiple times. Explicitly, we consider two states $\left| {\varphi _i } \right\rangle \left| {\varphi _j } \right\rangle$, we wish to determine whether $E_i  = E_j$. We have to perform PEA w.r.t the operator 
\begin{equation}
U = e^{ - iHt}  \otimes e^{ + iHt} \quad,
\end{equation}
which produces only the differences of two eigenvalues.

Suppose we normalize all our eigen-energy such that $0 < E_k < 1$ .We determine an energy window $\Delta = 2^{-a}$, for some integer $a$, the error $\varepsilon$ for a single run of the PEA is  bounded by
\begin{equation}
\varepsilon  < {{\Delta ^2 } \over { \left| {E_i  - E_j } \right|^2 }} \quad.
\end{equation}
If we apply PEA $k$ times with the same precision, then,
\begin{equation}
\varepsilon  \to \varepsilon ^k \quad, 
\end{equation}
which means that the errors for those energy changes $\left| {E_i  - E_j } \right|$ being greater than $\Delta$, $\left| {E_i  - E_j } \right| > \Delta$, would become exponentially small. The remaining problem is to deal with those energies change smaller than $\Delta$; this issue is related to the point (B) above. Let us call this new projector $\Lambda_1'$ which has resolution up to $\Delta$.

Consider the transformation described in Eq. (\ref{def_U_X}). Let us called the contribution from those ${\sigma _{ik} }$ where $\left| {E_i  - E_k } \right| < \Delta$ a ``leakage". It is because if those (nearly) degenerate eigenstates do not contribute to the resulting state, then the operator $ \Lambda_1' U_X^\dagger  U_Y \Lambda_1'$ would not transform any basis state $\left| i \right\rangle$ outside of the paired basis. In other words, ideally we want, for those cases where $\left| {E_i  - E_k } \right| < \Delta$,
\begin{equation}
\left\langle {\varphi _{k} } \right|K\left| {\tilde \varphi _i } \right\rangle = 0 \quad {\rm (ideal)}.
\end{equation}
Practically, this condition cannot be satisfied as typically $K\left| {\tilde \varphi _i } \right\rangle$ should have a board or almost continuous spectrum. However, we can argue that the sum of these contributions,
\begin{equation}
\eta  \equiv \sum\limits_k {'\left| {\left\langle {\varphi _k } \right|K\left| {\tilde \varphi _i } \right\rangle } \right|^2 }  \ll 1
\quad,
\end{equation}
could be made negligibly small for generic quantum systems where the Hamiltonian involves local interaction terms. Here the summation is over those $k$ where $\left| {E_i  - E_k } \right| < \Delta$. 

The reason is as follows: let us focus again on  Eq. (\ref{def_U_X}). Before applying $K$, note that both $\left| {\varphi _i } \right\rangle$ and $\left| {\tilde \varphi _i } \right\rangle$ have the eigen-energy $E_i$. We may expect, in the worst case scenarios, $\left| {\tilde \varphi _i } \right\rangle$ would contain a significant weight (i.e., sharply peaked) in those energy eigen-states close to $\left| {\varphi _i } \right\rangle$. In this case, the leakage in the degenerate subspace would be very bad. However, when we apply a ``kick" $K$ which does not preserve the symmetry of $\tilde H$, i.e., 
\begin{equation}
[ {K,\tilde H}] \ne 0 \quad ,
\end{equation}
then the change in energy 
\begin{equation}
\Omega  \equiv | {\left\langle {\tilde \varphi _i } \right|K\tilde HK\left| {\tilde \varphi _i } \right\rangle  - \left\langle {\tilde \varphi _i } \right|\tilde H\left| {\tilde \varphi _i } \right\rangle } |
 \quad 
\end{equation}
is typically of order $O(J)$, where $J$ is the typical size of the local terms. If we set $\Delta  \ll J$, then we expect that $
\eta  \ll 1$, and the correction to Eq. (\ref{Lambda_1}) and (\ref{Lambda_2}) is $O(\eta)$. The random kicks described in the previous section help spread out this effect (making the distribution more uniform). 

In short, as long as the distribution $\left| {\alpha _{k\tilde i} } \right|^2  \equiv \left| {\left\langle {\varphi _k } \right|K\left| {\tilde \varphi _i } \right\rangle } \right|^2$ is smoothly distributed over a range of energy which is much greater than the window $\Delta$ of the energy filter for $\Lambda_1'$, we should expect that the contribution coming from the ``leakage" can be made arbitrarily small, by decreasing $\Delta$. We leave a more quantitative analysis of this point for the future work.

\section{Appendix: Error analysis on the process of quantum simulated annealing}
Here we perform an error estimation for the process of quantum simulated annealing (QSA). First, the procedure of quantum simulated annealing starts with the infinite-temperature ($\beta=0$) state Eq. (\ref{int_state}), and end up at some finite-temperature ($\beta \ne 0$) state Eq. (\ref{coh_therm}). The inverse temperature $\beta$ is divided into uniform intervals 
\begin{equation}
\beta _j  \equiv \left( {j/d} \right)\beta \quad,
\end{equation}
where $j = 0,1,2,..,d$, of $d+1$. The coherent thermal states $| {\alpha_0^j }\rangle$ corresponding to the intermediate temperatures $\beta_j$ are prepared sequentially. The is made possible by the projective measurement which projects $| {\alpha_0^j }\rangle$ to $| {\alpha_0^{j+1} }\rangle$. The finally fidelity depends crucially on the overlap $| {\langle {\alpha _0^{j + 1} } | {\alpha _0^j } \rangle } |^2$ between these states. 

To estimate the overlap, note that  
\begin{equation}
| {\alpha _0^{j + 1} } \rangle  = \frac{1}{{\sqrt {\langle {\alpha _0^j } |e^{ - \Delta \beta H} | {\alpha _0^j } \rangle } }}e^{ - \Delta \beta H/2} | {\alpha _0^j } \rangle \quad,
\end{equation}
where $\Delta \beta  \equiv \beta /d$. The overlap, 
\begin{equation}
 | {\langle {\alpha _0^j } | {\alpha _0^{j + 1} } \rangle } |^2  =   \frac{{| {\langle {\alpha _0^j } |e^{ - \Delta \beta H/2} | {\alpha _0^j } \rangle } |^2 }}{{\langle {\alpha _0^j } |e^{ - \Delta \beta H} | {\alpha _0^j } \rangle }} \quad,
\end{equation}
is second-order in $\Delta \beta$, i.e.,
\begin{equation}
| {\langle {\alpha _0^j } | {\alpha _0^{j + 1} } \rangle } |^2  \approx   1 - O( {\Delta \beta ^2 \langle {H^2 } \rangle } ) \quad .
\end{equation}
This is analogous to the the quantum Zeno effect. In general, the energy fluctuation $\left\langle {H^2 } \right\rangle$ is smaller for thermal states of lower temperatures. Therefore, a potential improvement could be made by non-linear division of the $\beta_j$. Here we assume it bounded above,
\begin{equation}
\left\langle {H^2 } \right\rangle  \le \left\langle {H^2 } \right\rangle _{0} \quad.
\end{equation} 
For the whole process, the total error accumulates at each step to 
\begin{equation}
\epsilon  = d \times O( {\Delta \beta ^2 \langle {H^2 } \rangle _{0}} ) = O( {\beta^2 \langle {H^2 } \rangle_0 /d} ) \quad,
\end{equation}
as $\Delta \beta = \beta / d$. Hence, the total error can be made arbitrarily small by increasing $d$. In other words, to achieve any given accuracy $\epsilon$, one must perform at least 
\begin{equation}\label{app_d}
d = O( { \beta ^2 \langle {H^2 } \rangle_0 /\epsilon } )
\end{equation}
steps in the process of quantum simulated annealing. 

Next, for each step, the projective measurement 
\begin{equation}
\Pi _{k + 1}  = \sum_{k = 0}^{N - 1} {| {\alpha _k^{j + 1} } \rangle } \langle {\alpha _k^{j + 1} } |
\end{equation}
can be achieved by the phase estimation algorithm. For this purpose, an improved version is described in Ref. \cite{Wocjan2008} (Lemma 2). Here we summarize the result: to achieve an accuracy of $\epsilon_0$ , the number of application of the controlled-$W$ gates is 
\begin{equation}
O\left( {\log \left( {1/\epsilon _0 } \right)/\Delta _{\min } } \right) \quad,
\end{equation}
where $\Delta_{\min}$ is the minimun eigenvalue gap of $W$, see Eq. (\ref{Delta_min}). Here we should put $\epsilon_0 = \epsilon / d$, which equals $\epsilon ^2 /\beta ^2 \langle {H^2 } \rangle_0$ from Eq. (\ref{app_d}). Recall that $1/\Delta_{\min}$ is of order $O(1/\sqrt{\delta})$ from Eq. (\ref{Delta_min}), we conclude that the number of controlled-$W$ gate required is 
\begin{equation}
O\left( {\frac{{\beta ^2 \left\langle {H^2 } \right\rangle_0 }}{{\sqrt \delta  \epsilon }}\log \left( {\frac{{\beta ^2 \left\langle {H^2 } \right\rangle_0 }}{{\epsilon ^2 }}} \right)} \right) \quad.
\end{equation}
A quadratic speedup of $1/\sqrt{\delta}$ is achieved. This completes the error analysis.

\end{document}